\documentclass[sigconf,natbib=false]{acmart}
\copyrightyear{2024}
\acmYear{2024}
\setcopyright{rightsretained}
\acmConference[ICSE-SEIP '24]{46th International Conference on Software
Engineering: Software Engineering in Practice}{April 14--20, 2024}{Lisbon,
Portugal}
\acmBooktitle{46th International Conference on Software Engineering: Software
Engineering in Practice (ICSE-SEIP '24), April 14--20, 2024, Lisbon,
Portugal}
\acmDOI{10.1145/3639477.3639743}
\acmISBN{979-8-4007-0501-4/24/04}

\usepackage{biblatex}

\usepackage{amsmath,amsfonts}
\usepackage{algorithmic}
\usepackage{graphicx}
\usepackage{xcolor}
\usepackage{textcomp}
\usepackage{url}
\usepackage{hyperref}
\usepackage{multirow}
\usepackage{listings}
\DeclareMathAlphabet{\mathpzc}{OT1}{pzc}{m}{it}

\definecolor{codegreen}{rgb}{0,0.6,0}
\definecolor{codegray}{rgb}{0.5,0.5,0.5}
\definecolor{codepurple}{rgb}{0.58,0,0.82}
\definecolor{backcolour}{rgb}{0.95,0.95,0.92}

\lstdefinestyle{mystyle}{
    backgroundcolor=\color{backcolour},   
    commentstyle=\color{codegreen},
    keywordstyle=\color{magenta},
    numberstyle=\tiny\color{codegray},
    stringstyle=\color{codepurple},
    basicstyle=\ttfamily\footnotesize,
    breakatwhitespace=false,         
    breaklines=true,                 
    captionpos=b,                    
    keepspaces=true,                 
    numbers=left,                    
    numbersep=5pt,                  
    showspaces=false,                
    showstringspaces=false,
    showtabs=false,                  
    tabsize=2
}

\renewcommand{\paragraph}[1]{\textit{\textbf{#1}}}
\begin{document}

\title{LLM4PLC: Harnessing Large Language Models for Verifiable Programming of PLCs in Industrial Control Systems}

\author{Mohamad~Fakih$^1$, Rahul~Dharmaji$^{1*}$, Yasamin~Moghaddas$^{1*}$, Gustavo Quiros Araya$^2$,
Oluwatosin~Ogundare$^2$, and Mohammad Abdullah Al Faruque$^1$}
\affiliation{%
  \institution{$^1$Dept. of Electrical Engineering and Computer Science, University of California, Irvine, CA, USA \\
  {\text{$^2$Siemens Technology, Princeton, NJ, USA}}}
  \country{}}
  \affiliation{%
  \institution{\{mhfakih, rdharmaj, ymoghadd, alfaruqu\}@uci.edu \& \{gustavo.quiros, tosin.ogundare\}@siemens.com}
  \country{$^*$ These authors contributed equally to this project.}}

\renewcommand{\shortauthors}{Fakih, et al.}


\begin{abstract}
Although \textbf{Large Language Models (LLMs)} have established predominance in automated code generation, they are not devoid of shortcomings. The pertinent issues primarily relate to the absence of execution guarantees for generated code, a lack of explainability, and suboptimal support for essential but niche programming languages. State-of-the-art LLMs such as GPT-4 and LLaMa2 fail to produce valid programs for \textbf{Industrial Control Systems (ICS)} operated by \textbf{Programmable Logic Controllers (PLCs)}. We propose \textbf{LLM4PLC}, a user-guided iterative pipeline leveraging user feedback and external verification tools -- including grammar checkers, compilers and SMV verifiers -- to guide the LLM's generation. We further enhance the generation potential of LLM by employing Prompt Engineering and model fine-tuning through the creation and usage of \textbf{LoRAs}. We validate this system using a \textbf{FischerTechnik Manufacturing TestBed (MFTB)}, illustrating how LLMs can evolve from generating structurally-flawed code to producing \textbf{verifiably correct programs} for industrial applications. We run a complete test suite on \textbf{GPT-3.5, GPT-4, Code Llama-7B, a fine-tuned Code Llama-7B model, Code Llama-34B, and a fine-tuned Code Llama-34B model}. The proposed pipeline improved the generation success rate from 47\% to 72\%, and the Survey-of-Experts code quality from 2.25/10 to 7.75/10. 

\textbf{To promote open research, we share the complete experimental setup, the LLM Fine-Tuning Weights, and the video demonstrations of the different programs on our dedicated webpage\footnote{https://sites.google.com/uci.edu/llm4plc/home}.}

\end{abstract}

\maketitle

\section{Introduction}

\begin{figure}[h!]
  \includegraphics{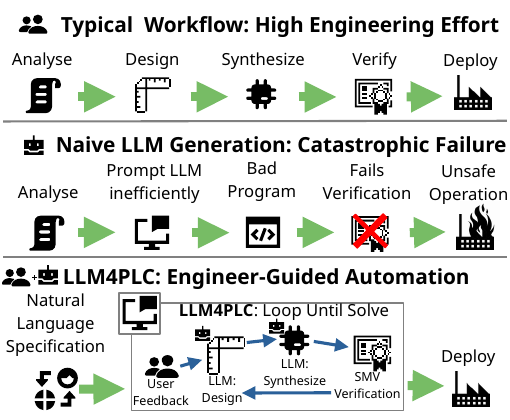}
  \caption{Our proposed LLM-augmented workflow automates the high-effort stages of the PLC programming methodology}
  \label{fig:teaser}
\end{figure}

Programmable Logic Controllers (PLCs) are indispensable in the landscape of Industrial Automation -- a market valued at $\$180$~billion US Dollars in  2022~\cite{ISA} -- and these controllers drive essential infrastructure and industry such as oil pipelines \cite{rashad2022plc}, electric grids\cite{grid}, manufacturing sites \cite{werner2020supporting}, and nuclear power plants \cite{first2022diversity}. PLCs are domain-specific real-time computers, integrating an \textit{"Input-Compute-Output"} execution loop and running specialized programs created with one of five programming paradigms standardized under IEC 61131-3 \cite{iec}. Out of these five approaches, only \textbf{Structured Text (ST)} resembles conventional programming languages in regards to its syntax and structure. This property allows for automated code generation targeting the ST language using state-of-the-art techniques. Moreover, the usage of formal verification schemes for IEC 61131-3 programs \cite{ovatman2016overview} enables generated code to meet strict safety, complexity, and timing requirements.

Software in critical infrastructure and machinery are required to operate within a narrow safety margin and typically \textbf{necessitate extensive testing and verification}. In the typical project lifecycle, engineers and domain experts extensively analyze and design potential solutions before any programming effort is made, followed by dedicated synthesis and verification steps before deployment \cite{SoftEngineeringMethods}. A visualization of this workflow is shown in Figure~\ref{fig:teaser}. Therefore, the governance of PLC control programs by \textbf{strict guidelines and requirements}, adds complexity to engineering tasks, resulting in hundreds of extra hours of expert-level effort, often even requiring Reverse Engineering \cite{cfg2vec} to recover the initial intent of the programmer. The primary goal of our proposed pipeline is \textbf{expediting the engineering effort} by offloading a sizeable majority of PLC-related problem-solving tasks to a dedicated LLM-Agent that assists engineers in their jobs. Our approach contrasts sharply with existing automated programming approaches, where the engineer is required to create the model design, synthesize code, and verify their solution manually.

Although some techniques exist to automate the synthesis of IEC-61131-3 PLC programs \cite{hao2019test, vogel2005uml} given a specification and synthesis paradigm -- such as Linear Temporal Logic \cite{kuzmin2013ltl}, or novel software-implemented frameworks such as MODI \cite{cai2022plc} -- the engineering challenge of combining all parts of the PLC programming pipeline into a single unified model remains.

Recent developments in Large Language Models (LLMs) offer an alternative to legacy automation methods. However, given the irregularities in LLM code generation \cite{siddiq2023code}, naive use of LLMs in the Engineering Workflow -- signified by inefficient prompting and blind execution of unverified output code -- leads to unsafe operation \cite{pierce2021repair}, as showcased in Figure~\ref{fig:teaser}. Yet, foundational models such as GPT-4 \cite{gpt4} and LLama 2 \cite{llama2} are challenging traditional approaches to automation and programming. Especially noteworthy is these models' instructional (i.e., "chat-instruct") capabilities, which allow for dynamic prompting based on a conversational input paradigm, opening the door for automated feedback mechanisms. Additionally, with the application of Parameter-Efficient Fine-Tuning (PEFT), Low-Rank Adaptations \cite{hu2021lora} have made domain-specific training easier and significantly reduced compute and data requirements for these tasks \cite{liu2022peft, xiong2023doctor}. Lastly, the LLM research community has pioneered "prompt-engineering" -- a practice of optimizing prompts resulting in more accurate and relevant LLM responses \cite{spasic2023prompt, strobelt2023prompt}. These advances offer a unique opportunity for combination with LLM-based code generation techniques for Industrial Control Systems, forming the basis of our automation pipeline.

We introduce LLM4PLC, an automated pipeline that integrates Large Language Models (LLMs) with industry-standard PLC systems, employing automated verifiers and optional human feedback to ensure safe and efficient code deployment. After receiving an initial Natural Language Specification for the specified PLC system, the LLM Agent enters an automated iterative loop: generating a design schematic, synthesizing Structured Text (ST) code, and undergoing a sequential verification process that includes syntax checking \cite{nucleron2017}  and model checking via the NuXmv software suite \cite{cavada2014nuxmv}. On a verification-stage success, the code can be immediately deployed; otherwise, errors from the verification stage are fed back to the LLM Agent for refinement of the erroneous ST code, with the option for human intervention during this process. This workflow is illustrated in Figure~\ref{fig:teaser}. Therefore, we summarize our key contributions as follows:
\begin{itemize}
    \item Our work proposes and implements an automated language-driven system to verifiably program PLC devices from natural language descriptions of industrial plants.
    \item To the best of our knowledge, our work is the first to propose augmenting Large Language Models with automated external code verifiers to converge toward a solution iteratively.
    \item We present a detailed study of the generation potential of prevalent LLM models: GPT-3.5, GPT-4, Code Llama, and our fine-tuned Code Llama. We measure each configuration's generation success rate and average expert-appointed score.
\end{itemize}
The remainder of this paper is organized as follows: Section~\ref{sec:related} offers a literature review to ground our research within the existing academic landscape. Section~\ref{sec:background}, creates a framework for the foundations of PLC systems and identifies the issues our solution aims to address in LLM generation. This section also addresses background knowledge for other stages of our pipeline, including  LoRA creation, syntax checking, and formal verification. In Section~\ref{sec:method}, we expand on the methodology behind LLM4PLC, detailing the design, implementation, and verification stages of our automated pipeline. Section~\ref{sec:setup} and Section~\ref{sec:results} presents the experimental setup and results, showcasing the efficacy and reliability of our approach through quantitative evaluations. Finally, Sections~\ref{sec:discussion}, \ref{sec:industry}, and~\ref{sec:conclusion} discuss the broader implications of our findings, potential limitations, future directions, and conclude the paper.

\section{Related Works}
\label{sec:related}
Our work is primarily related to two fields of prior research: automated PLC programming and LLM Augmentation.

\paragraph{Automated PLC Code Generation} has previously been explored by methods that do not utilize LLMs. In \cite{schumacher2013tool} the authors present a technique to translate GRAFCET -- a graphical modeling language -- into IEC 61131-3. Their editor binds GRAFCET elements to user-specific shapes with the ability to customize the behavior of each shape. Then, the graph is translated into an internal object model that is parsable by the GRAFCET toolchain, yielding code that might be executed in parallel \cite{parallel}. The designer is required to provide a reference graphical solution, in contrast to our LLM-driven approach, which requires a set of natural-language instructions that define the desired operation of the machine.

Alternatively, previous works have developed an organization-level automation assistant. \cite{puttonen2012semantics} proposes using a web service agent to assist production processes. Their methodology requires that each machine operating on the factory floor is be accessible via a web-interface implementing a well-defined API. Using OWL-S \cite{owl-s} and SPARQL \cite{sparql}, a world model is built according to the organization of web interfaces, and the internal states of the system are frequently updated due to Service Monitor invocations. OWL-S provides prescriptive commands to processes based on production goals created by SPARQL. Their tool only abstracts away the physical component of the program automation -- a system design without physical access to the machines will still require that design and engineering tasks be handled by an adept engineer.

\paragraph{The use of LLMs in Industrial Automation} has already been attempted in previous works. The team behind \cite{deppe2022ai} created an intelligent assistant to assist in tasks such as process execution and troubleshooting. First, a knowledge graph based on information gleaned from user manuals is created. Experts then enhance the graph with their domain-specific knowledge. For each query, the information retrieval system extracts relevant text from the graph database, achieved in part by representing text passages as dense vectors using language models such as BERT or GPT. A limitation of this approach is that language models typically fail to return domain-specific information. As a result, the researchers trained a separate model to alleviate this shortcoming, and then combined their results with a pre-trained language model. Afterwards, the selected text passages are ranked by order of helpfulness, desirability, and relevancy. While this approach is helpful for maintenance and operations, it does not automate any part of the design and implementation phases.

\paragraph{LLM Finetuning for Code Review Automation} has also been attempted by \cite{lu2023llama} through the use of LoRA-based methods to improve the Llama LLM. The authors' approach uses a \\ LoRA-Augmented LLM as an automated Code Review agent and forwards its feedback into the code-generation agent to enable better quality code output. Their findings show that LoRA is the preferred method for fine-tuning code generation in this context. Supplying stricter feedback in this process has also been explored in CodeRL \cite{le2022coderl}, where the LLM is finetuned using Reinforcement Learning techniques by leveraging the pass rate on generated unit tests and compilers as a reward function. These two methods require a paired dataset of candidate code and automated feedback, which is difficult to acquire and may introduce biases into the framework. Rather than focusing on refinement in the training stage, our approach explores iterative refinement in the inference stage.

\paragraph{Dataset Preparation} is also a crucial stage for code generation. Foundational Large-Language Models such as GPT-4 or Llama 2 are trained on a large and varied corpus of text \cite{llama2, gpt4}. Recent efforts have focused on improving code generation or completion from LLMs through fine-tuning these models on specific input data. For instance, Thakur et al. focused on automatically generating hardware description language (specifically Verilog) using LLMs \cite{thakur2023verigen}. They fine-tuned various LLMs on a large Verilog corpus, which they assembled from project files on Github and numerous academic textbooks. Finetuning involved sharding the optimizer states across GPUs instead of using a quantized model or LoRAs. Other attempts at finetuning for specific output domains include natural language response \cite{wei2021zeroshot, liu2020roberta} and the esoteric language Hansl \cite{tarassow2023hansl}.

The authors in \cite{guo2022unixcoder} present UniXCoder, an ML model incorporating \textit{Abstract Syntax Trees (ASTs)} and comments into the generation scheme to create better output. Specifically, they are extracting the AST during the generation process so that the ML model can consider it while generating code. They also use comments as a source of guidance, as they are a helpful tool for understanding the underlying details and assumptions of a function, program, or algorithm. Similar to our approach, the authors create separate code completion and code generation datasets in order to validate their claims.

Other research groups have opted to build externally augmented language models -- Tang et al. target software development languages by developing a database-equipped language model for domain adaptive code completion without fine-tuning \cite{tang2023domain}. They retrieve information from a database separate from the language model, and therefore avoid excessive reliance on the weights of the language model. The next step is to use Bayesian inference for interpolating between the results of this database and the language model. Experimental results show improved performance of both CodeGPT \cite{lu2021codexglue} and UniXCoder \cite{guo2022unixcoder}. Their method, while useful, does negatively impact completion speed. Ultimately, their conclusion indicated that fine-tuning was superior in terms of accuracy and code quality.

\paragraph{Code Verification} is an important step for evaluating the LLM. There have been various methods for code verification in the literature, including control flow analysis, dynamic symbolic execution, and model checking.

\cite{prahofer2012opportunities} introduces the necessary software to perform static analysis of PLC programs. They cover rule-based approaches, syntax checking, and other techniques commonly accepted by the scientific community. \cite{prahofer2016static} describes more complex techniques, such as generating an AST and performing control flow analysis to enable the creation of detailed insights regarding the logic flow of a PLC program. In our method, we build upon accepted static analysis knowledge in our verification pipeline, ultimately using this pipeline to verify the LLM-generated code.

Other works, such as \cite{he2021automated}, employ dynamic symbolic execution to generate test cases given a PLC code sample. Incorporating dynamic symbolic verification into a verification pipeline simplifies the process of proving functional correctness. Automating test generation removes human errors induced via manual test generation, such as missing test cases or redundant branch checking.

Model checking is a technique for verifying that the specifications defined for a model are met. \cite{ovatman2016overview} covers model checking as a verification scheme for PLC programs. The authors include SMV-based checking as a means to validate PLC software. We draw ideas from this paper regarding the usage of an SMV toolchain to perform formal model-based verification.

\vspace{-0.5em}
\section{Background}
\label{sec:background}

The methodology adopted to develop our proposed pipeline builds upon the existing wisdom on PLC Software Engineering, Formal Verification Methods, and State-of-the-Art techniques in LLM prompting as well as Parameter Efficient Fine-Tuning (PEFT) using Low-Rank Adaptations (LoRAs). In this section, we provide the reader with the necessary prerequisite knowledge of each domain, while also setting up our motivation for the design choices adopted in the proposed method. First, we showcase the \textit{prevailing approaches for efficiently querying LLMs} and the \textit{fine-tuning methods for injecting knowledge into LLMs}, and then we delve into the \textit{formal verification techniques used extensively in PLC programming}. 

\subsection{Large Language Models}
Large language models (LLMs) leverage the attention mechanism in the Transformer architecture\cite{vaswani2017attention} to model sequences of increasing lengths. At the core of the Transformer model lies the self-attention mechanism, which computes attention scores for each pair of tokens in a sequence, allowing for long-range dependencies and relationships between tokens across long distances within the sequence.

All major LLMs use a Next-Token-Prediction (NTP) scheme to generate one token at a time. Formally, given a dictionary of possible tokens $\mathbb{T}=\left\{T_1, T_2, ..., T_N\right\}$, and a sequence of tokens $S =\left\{T_{s_1}, T_{s_2}, ... T_{s_m}\right\}\in \mathbb{T}^{m}$, an LLM model $\mathcal{M}(.) \in R^{N}$ attempts to model the likelihood of the next token $T_{s_{m+1}}$ given:
\begin{equation}
    \mathbb{P}\left(s_{m+1} = i | S \right) = \mathcal{M}\left(S\right)_i
\end{equation}
Therefore, the probability of sampling a continuation sequence $\hat{S} = \left\{T_{s_{m+1}}, T_{s_{m+2}}, ..., T_{s_{m+M}}\right\}$ is expressed as:
\begin{equation}
    \mathbb{P}\left( \hat{S} | S \right) = \prod_{i=0}^{M} \mathcal{M}\left( T_{s_1}, T_{s_2}, ..., T_{s_{m+i}} \right)_{s_{m+i+1}}
\end{equation}

LLMs have gained substantial traction since the release of OpenAI's models, GPT-2 and GPT-3. At its release, GPT-2 stood out for its large size of 1.5 billion parameters \cite{radford2019language}. Almost a year and a half later, GPT-3 came out with a staggering 175 billion parameter model. Larger models benefit from enhanced learning and the ability to store more information about complex relationships in data. While the results of OpenAI's models are remarkable, the model weights are not open-source, therefore eliminating the possibility of LLM fine-tuning\footnote{Between the initial writing of this paper and its acceptance, OpenAI introduced fine-tuning support for their language models} and creating a data privacy issue.

One of the first open-source models was Llama, released by Meta \cite{touvron2302llama}. Contemporary results show that the 70B parameter Llama model performs comparably to GPT-3.5 on most benchmarks \cite{touvron2302llama}. The most recent Llama release and the one most pertinent to our research is Code Llama \cite{roziere2023code}, whose base model is Llama 2 fine-tuned on code-specific datasets \cite{roziere2023code}.

\subsubsection{Prompt Engineering}

Once the LLM architecture has been selected, one can begin prompting the LLM as part of training and inference. Prompt engineering refers to creating prompts instructing the LLM to produce a desired response. Prompts are meant to provide context, thereby facilitating LLM response generation. For example, in writing code for a specific language, one could create a prompt that provides context by including an example program written in the language and then asking the LLM to fix a piece of code. This is exactly how our approach goes about prompting the LLM during training. During inference, the prompt is simply a command that asks the LLM to complete unfinished code.

Formally, a prompt $\mathcal{P}$ is a sequence of tokens 
$\left\{ T_{p_1}, T_{p_2}, ..., T_{p_{K}}\right\}$ that is inserted as a prefix for any sequence $S$. Different prompts naturally lead to varied degrees of success, with "self-guidance" prompts achieving the best results \cite{prompt_chain}. Self-guidance prompts decompose a thought process into steps. In fact, we leverage self-guidance when we attempt compilation of the code outputted by the LLM and use any resulting errors as feedback to the LLM so that it can make the necessary corrections. 

\subsubsection{Parameter Efficient FineTuning (PEFT)}

PEFT aids in adapting an LLM to a particular task. LLMs typically have billions of parameters and training them on new tasks can be computationally demanding and time-consuming \cite{naveed2023llm,sanyal2023llm,tornede2023llm}. PEFT encompasses techniques that improve model performance on a specific predefined task without compromising resource efficiency or training duration. Of the many approaches to PEFT, Low-rank Adaptation (LoRA) is a popular, performant one, and our pipeline heavily incorporates this technique.

\begin{figure}[h!]
    \centering
    \includegraphics[width=\linewidth]{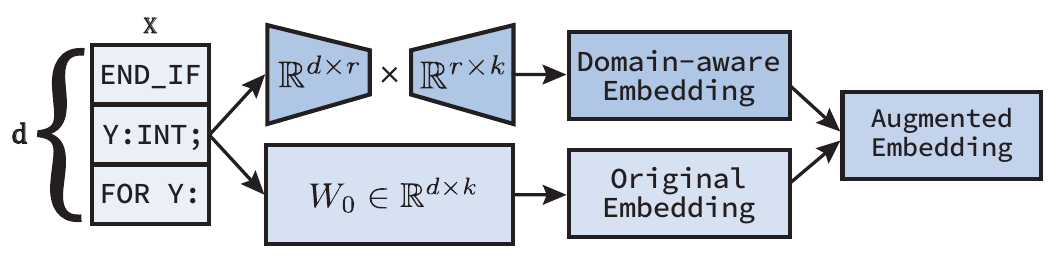}
    \caption{LoRAs create domain-specific embeddings that are then aggregated into the original knowledge base}
    \label{fig:lora}
\end{figure}

LoRAs help address the common issue of over-fitting and catastrophic forgetting \cite{coop2013,yang2022peft}. The LoRA technique is applied to reduce the memory that is used up by the update weights, $\Delta$\textit{W}. This process involves low-rank decomposition of the update weight matrices \cite{hu2021lora}
as can be seen in Equation \ref{LoRA_equation}, where 
$W_0 \in \mathbb{R}^{dxk}$,
$B \in \mathbb{R}^{dxr}$,
$A \in \mathbb{R}^{rxk}$,
and $r \ll min(d,k)$. The process is visualized in Figure~\ref{fig:lora}, where the tokens are passed as an input, \textit{x} with dimension \textit{d}.

\begin{equation} 
\label{LoRA_equation}
    W_0x + {\Delta}Wx = W_0x + \alpha * BAx
\end{equation}

Instead of using a very large matrix, the update matrices are broken down into two smaller matrices through low-rank decomposition. Our work further exploits the benefits of this technique by training multiple LoRAs in parallel. In essence, the additional domain-specific knowledge is injected into the auxiliary network, offering augmented problem-solving capabilities without a degradation in latency. The size of these auxiliary passes is controlled by the rank $r$, a tunable parameter that controls the size of the LoRA. This parameter controls both the model's efficacy at incorporating new knowledge as well as its training time and required dataset size. Lastly, the LoRA injection is controlled by a strength scale $\alpha$, which dictates the extent to which the auxiliary network influences the primary model. Then, $\alpha$ serves as a tunable hyperparameter, allowing for a calibrated trade-off between domain-specific expertise and generalization performance. By adjusting $\alpha$, we can fine-tune the model's reliance on the LoRA, effectively balancing the incorporation of specialized knowledge against the risk of overfitting to a particular domain.

\subsection{Model-Based Design}
Model-Based Design (MBD) is an engineering paradigm that leverages mathematical and graphical modeling to facilitate the analysis, implementation, and simulation of complex systems. Originating from control engineering and systems theory \cite{mosterman2007mbd}, it has been successfully applied across diverse domains, including automotive\cite{mudhivarthi2023mbd}, aerospace \cite{bachelor2020mbd}, cyber-physical systems \cite{jensen2011mbd}, and industrial automation \cite{streit2018mbd}.

\begin{figure}[h!]
  \centering
  \includegraphics[width=\linewidth]{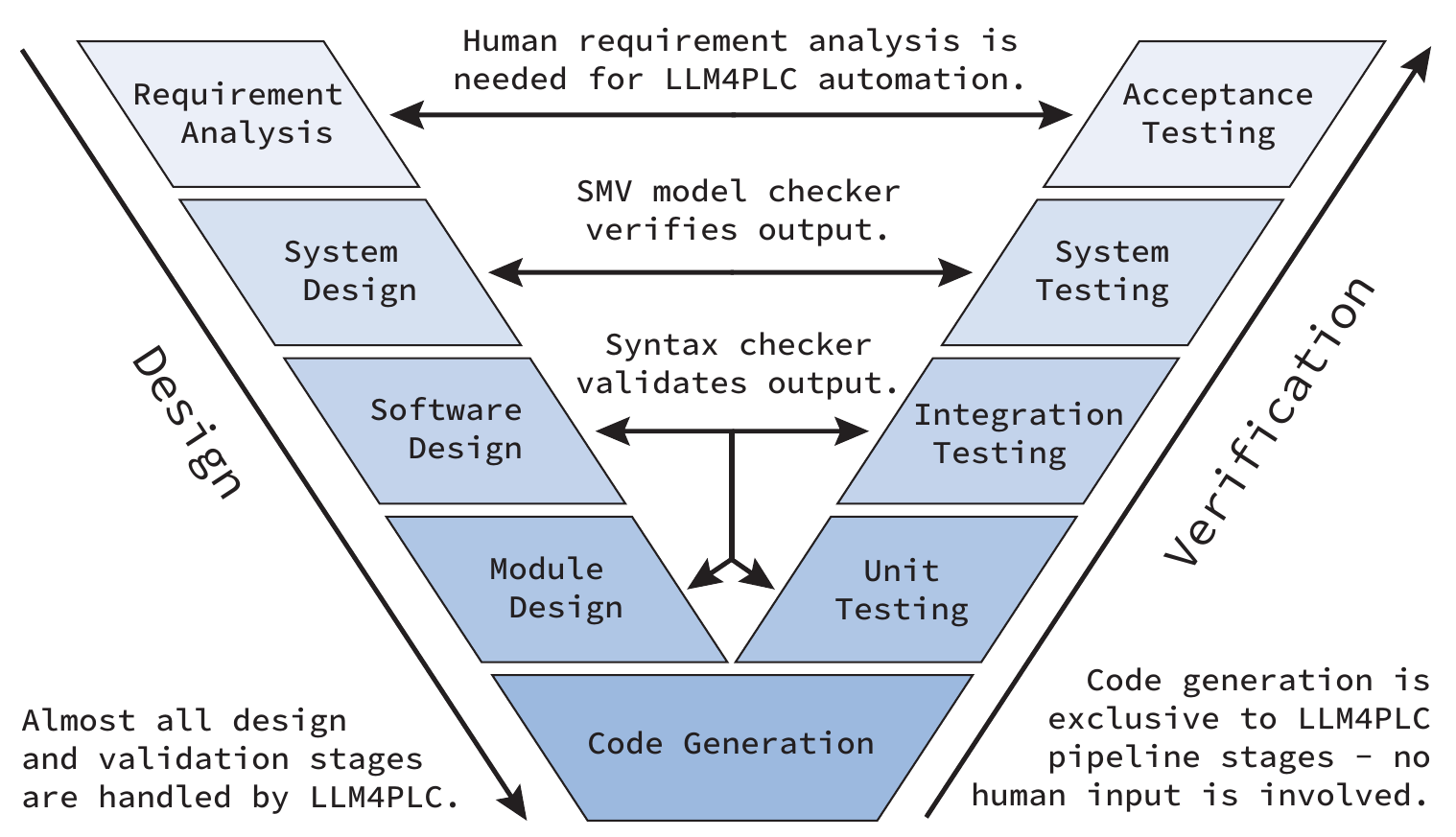}
  \caption{An example of a Model-Based Design process}
  \label{fig:mbd}
\end{figure}

In traditional design methodologies, each development stage -- requirements engineering, architecture design, implementation, and verification -- is often isolated, requiring manual and often error-prone intervention to transition between stages. MBD, on the other hand, emphasizes an integrated framework where predefined models serve as a formal and comprehensive representation of the system, eliminating these transition-induced errors. These stages are standardized in \cite{mosterman2007mbd}. An example of MBD can be seen in Figure~\ref{fig:mbd}.

Lastly, verification and validation processes are tightly integrated into the typical MBD design flow. Because the model serves as the golden reference for the system, it can be used to rigorously test the final implementation, ensuring that it meets the agreed-upon requirements and constraints.

Model-based design is a holistic approach to system development that offers significant efficiency, reliability, and maintainability benefits. Its emphasis on early-stage simulation and analysis enables proactive problem-solving, making it an increasingly essential technique for designing and implementing complex systems. The typical engineering workflow previously presented in Figure \ref{fig:teaser} adheres to this paradigm. LLM4PLC uses these successive stages as guidelines for querying the LLM as we present in section~\ref{sec:method}.

\subsection{Syntax Checkers and Formal Verification}

Without a methodology in which human-generated or LLM-generated code can be evaluated for safety, completeness, and accuracy, a number of bugs and other undesirable behavior may persist in the code. In environments such as nuclear power plants, or satellite control systems, correctness is not only desirable but wholly necessary for operation \cite{first2022diversity}.  It is well known that LLMs often produce code that does not conform to language specifications \cite{siddiq2023code}. There are several ways in which software can be checked for these deficiencies, many of which vary in complexity and effectiveness. We are primarily concerned with syntax checkers and compilers, which provide a first step towards ensuring the correct operation of the program. Then, syntax checkers and compilers help alleviate this issue by providing feedback on LLM-generated code.

\subsubsection{Syntax Checkers}

The first step in transforming candidate code to a working program is to check that it conforms to the standards of the programming language used. Without this step, the code will not compile, and remedial action is required to be taken. Automated approaches to fix syntax errors exist \cite{ahmed2023synshine}, but integrating syntax checkers into an LLM code generation pipeline enables the repair of errors previously unaccounted for. Using these tools over multiple cycles of the LLM4PLC pipelines provides a deeper insight into code deficiencies, and better prepares the LLM to take corrective action.

\subsubsection{Formal Verification}

Formal verification of programs and algorithms through Symbolic Model Checking is an essential step toward deploying PLC code in hazardous environments. These tools take in the candidate code as well as strict constraints on the operation - for example, the upper limit of temperature, or . Approaches using interactive theorem provers \cite{blech2013on}, or Symbolic Model Checking \cite{blech2013on} have become widely used for this purpose. The focus of LLM4PLC is on the usage of these tools in order to formally verify LLM-generated code using an SMV model generated from a plant specification document. Then, the generated PLC code is verified using the SMV model as the reference. 

\section{LLM4PLC Methodology}
\label{sec:method}

LLM4PLC is a user-guided iterative pipeline designed to help LLMs generate code for Industrial PLCs. We aim to address the limitations of current state-of-the-art Large Language Models (LLMs) in this domain. Our pipeline integrates user feedback loops and incorporates a suite of external verification tools: grammar checkers and a nuXmv verifier. The pipeline is optimized via Prompt Engineering and model fine-tuning mechanisms utilizing LoRAs.
\begin{figure}[h!]
    \centering
    \includegraphics[width=\linewidth]{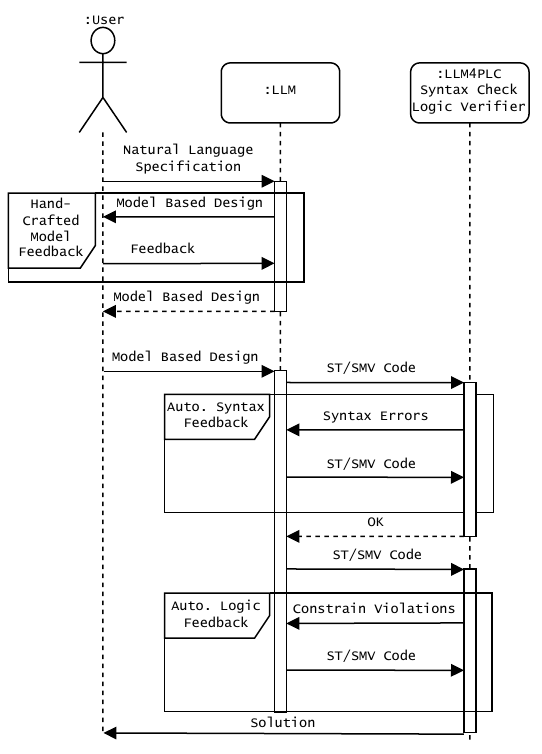}
    \caption{UML Representation of proposed engineering pipeline}
    \label{fig:uml}
\end{figure}

\begin{figure*}[!ht]
    \centering
    \includegraphics[width=\linewidth]{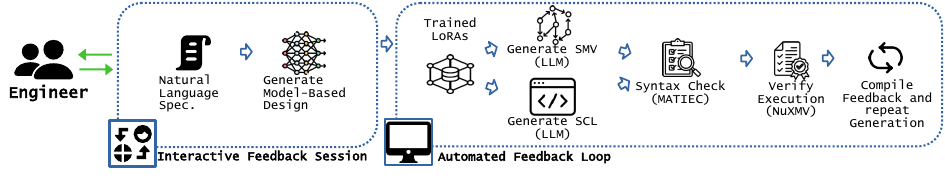}
    \caption{Summary of all blocks used in LLM4PLC}
    \label{fig:doublecol}
\end{figure*}

The approach of the implementation follows a top-down view of the typical workflow in industrial settings where the engineer's effort is showcased in a representative UML diagram in Figure~\ref{fig:uml}. 
Additionally, for better clarity on how blocks are interconnected, we present all modules used in LLM4PLC in Figure~\ref{fig:doublecol}. We discuss the flow in the following sections.

\subsection{Model-Based Design}
\paragraph{Model-Based Design (MBD)} in the context of industrial PLC programming introduces a structured, systematic approach that enhances both the efficiency of the development process and the reliability of the resultant code. By following task-specific prompt guidelines for planning, the LLMs can sift through given specifications and requirements to create a comprehensive plan for the subsequent pipeline stages. The first step in LLM4PLC is to generate complete function and block declarations and their corresponding signatures, serving as an executable blueprint that ensures alignment between the development process and the defined natural language specifications. Moreover, the detailed planning phase can highlight any ambiguities in user requirements, providing an opportunity for prompt clarification and thereby minimizing the risk of deviations or errors in the subsequent code implementation phase. An example MBD plan generated by our pipeline exists in our dedicated website page\footnote{https://sites.google.com/uci.edu/llm4plc/home}.

\paragraph{Finite State Machines (FSMs)} provide a significant advantage for our development methodology. Explicit planning around FSM states not only allows for a clear roadmap but also optimizes the PLC scan cycle. Because PLC logic does not follow the traditional loop-based execution model, using a state variable to track the system state is crucial. Each state in the FSM is responsible for a single operation, making the execution predictable and easier to debug. In the MBD prompt used alongside the Natural Language Specification, we constrain the LLM to follow an FSM design for the design solution.

\subsection{Syntax Checkers}

Using LoRAs, we transform the MBD plan into its associated Structured Text representation in Siemens' Structured Control Language. Then we use an open-source IEC 61131-3 Structured Text compiler in order to perform syntax checking of LLM-generated code. Specifically, we use MATIEC \cite{nucleron2017} to search for syntax errors in the generated code. If any error is detected, the output of MATIEC is then fed into our pipeline in order to create a 'correction prompt' for the LLM in the next stage of the pipeline. We feed only one compiler error and its associated generated prompt per cycle for several reasons. Firstly, we want a targeted fix from the LLM for each error, rather than continuously prompting to fix multiple errors at once. Moreover, for each compiler error, subsequent errors can be dependent on the original error. Take for instance a missing semicolon which directly causes another compiler error on a subsequent line. By feeding only one error and correction prompt at a time, we work to minimize the total number of prompts and pipeline iterations needed. Note that this would also allow for multiple errors to be fixed in one iteration cycle (i.e. when one error is wholly caused as a result of a preceding error).

\subsection{Formal Verifiers}
Upon generating compilable PLC code through our pipeline, the next step requires verification using nuXmv, a symbolic model checker based on the SMV paradigm \cite{cavada2014nuxmv}. This verification process is crucial to verify that the produced code adheres not only to syntactical standards but also to functional and safety requirements intrinsic to industrial automation scenarios.

To conduct this verification, we translate the constraints from Natural Language to SMV using a feedback loop similar to the one used for ST. The provided plant specification outlines the behavior, constraints, and requirements of the industrial process the PLC code is intended to control. The SMV specification file, on the other hand, encapsulates the formal properties and conditions that the PLC code must satisfy. We use these two elements in conjunction with nuXmv as a means to perform formal verification on the PLC code. This approach ensures that the generated code is not only compilable but also reliable and safe for deployment in a real-world industrial setting. Through this verification mechanism, any discrepancies between the intended and actual behavior of the PLC code can be promptly identified and rectified before deployment, enhancing the robustness and credibility of our pipeline.

\section{Experimental Setup}
\label{sec:setup}

For the purposes of our study, we target GPT-3, GPT-4, Code Llama 7B, and Code Llama 34B, motivated by several considerations towards a comprehensive evaluation. GPT-3 \cite{ye2023gpt} and GPT-4 \cite{gpt4} represent state-of-the-art general-purpose language models, serving as effective benchmarks for general text-to-code translation tasks. Code Llama 7B and Code Llama 34B \cite{touvron2302llama} are specifically designed for code generation: Their respective 7-billion and 34-billion parameter counts offer a gradient of computational complexity, enabling the investigation of the trade-offs between performance and computational resources. 

\subsection{Prompting}
Two prompt types, \textit{Zero-Shot} and \textit{One-Shot}, were employed to generate outputs from the LLMs. The One-Shot prompt incorporates a representative sampling of Structured Text syntax and code elements, thereby providing contextual cues for improved code generation. In contrast, the Zero-Shot prompt simply requests code generation without any contextual guidance. For each model we assess its performance against both prompt types. 

\subsection{LLM Fine-Tuning}
\label{sec:finetune}

We fine-tune the Code Llama 7B and Code Llama 34B through the creation of Low-Rank Adaptions (LoRAs). GPT-3.5 and GPT-4 do not provide an interface for the training of LoRAs, so we use these models in their default configurations. Then, we leverage this knowledge to train a set of four LoRAs for both code completion and code fixing tasks -- two for each of Code Llama 7B and Code Llama 34B. We deploy training and inference on a compute node equipped with an Nvidia A100 as well as 128GB of main system memory.

\subsection{Dataset}

We generate a training, validation, and testing dataset from the OSCAT IEC 61131-3 Library \cite{OSCAT2023}. Specifically, we run automated tests to cull deficient ST files in the OSCAT dataset, then create three separate datasets to test our pipeline's capabilities extensively.

\subsubsection{Automated Tests}

We attempted to compile each ST file in the aforementioned dataset to validate our testing data and discarded all files that did not successfully compile. MATIEC is chosen as the ST compiler for this task. This left us with 636 viable ST files to create our dataset from. We use 596 samples for Training and 40 samples for testing (95/5 Train/Test split).

\subsubsection{Dataset Generation}
\label{sec:dataset}
We derive three datasets from the original OSCAT Dataset: (1) Generation, (2) Completion, and (3) Fixing. The generation dataset is simply the OSCAT dataset after culling all non-compilable files. We build the Completion dataset by randomly truncating files in the Generation Dataset. This allows us to simulate the code-completion abilities of the LLM and test the ability of our pipeline to aid in the code-completion task. The Fixing dataset is created by removing random lines from the Generation dataset until the resulting ST file is no longer compilable. This dataset aims to test the LLM's ability to synthesize solutions to specific syntax errors within the code. Once again, we use MATIEC in order to verify the compilation status of each file during this process.

\subsubsection{LoRA Training}
As described previously, we finetune a pair of LoRAs for each model: one for completion and one for fixing. Section \ref{sec:background} presents several parameters relevant for training LoRAs. We choose rank $r=64$, strength scale $\alpha = 128$, and $Batch\_Size=256$ and we run our training procedure over 5 epochs. These parameter choices were made according to prevalent wisdom in the LLM community as well as our team's experience in training and deploying such models.

\subsection{Metrics}
\label{sec:metrics}

Our metrics involving the experimental dataset include pass rate, compiler error count, and a human evaluation of code quality. Pass rate is evaluated using the pass@k metric, as outlined in \cite{chen2021evaluating}. For each of the 40 test files, a single output is generated (k=1), yielding statistically significant results. Finally, we compare the amount of Engineer-Hours required for each of the configurations.

\paragraph{The pass rate}, assessed via the pass@k metric, serves as an indicator of the method's accuracy in code generation. A high pass rate indicates the procedure's reliability in producing syntactically and semantically correct code.

\paragraph{The compiler error rate} offers insight into the robustness of our approach. A lower rate signifies a reduced likelihood of generating syntactically flawed code, highlighting the method's precision in adhering to language-specific rules. We track the number of detected syntax errors per generated files and compute the average errors per file.

The pass rate and error rates are not individually indicative of success. Since LLM4PLC rejects all code that contains errors, an increase in error count does not necessarily imply a failure of the pipeline. This metric serves as an auxiliary measure to examine after the pass rate has been considered. For example, if a pipeline stage has a higher pass rate but also an increase in compiler error count, this is still beneficial as only the successful iterations (i.e. those with zero errors) are forwarded to the verification stage of the pipeline. Additionally, if a pipeline stage has a lower pass rate but a low compiler error count it does not indicate success, since success is firstly measured by compiler pass rate. 

\paragraph {The human code quality evaluation} provide a nuanced evaluation of the generated code's readability and maintainability. High scores in this dimension underscore the practicality of our method, emphasizing its potential for seamless integration into existing human-driven development processes.

\subsection{Testbed}

Our lab has deployed the FischerTechnik Manufacturing Testbed (MFTB)\cite{FischerTechnik2023} -- an integrated platform that simulates a miniaturized version of common manufacturing processes. The MFTB serves as a sophisticated test environment for studying and validating various aspects of automation, digital control systems, and operational efficiency in a manufacturing setting. The testbed is a complex cyber-physical system integrating various inputs and different types of outputs as follows:

\begin{itemize}
    \item \textbf{Digital Inputs:} 22
    \item \textbf{Analog Inputs (0-10V DC):} 1
    \item \textbf{Fast Counting Inputs:} 10 (for direction detection)
    \item \textbf{Outputs (24V):} 35
\end{itemize}

\begin{figure}[h!]
    \centering
    \includegraphics[width=\linewidth]{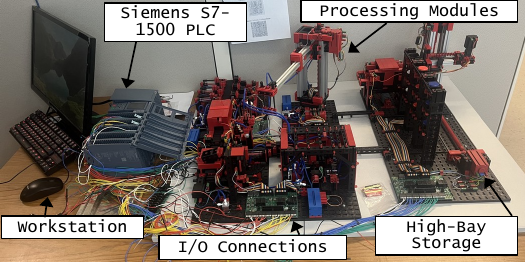}
    \caption{The FischerTechnik Manufacturing Testbed (MFTB) deployment in our lab}
    \label{fig:mftb_neutral}
\end{figure}

\paragraph{Functional Modules} in the MFTB serve four distinct purposes, and are summarized as follows:
\begin{enumerate}
    \item Sorting Line With Color Detection
    \item Multi Processing Station With Oven
    \item Automated High-Bay Warehouse
    \item Vacuum Gripper Robot
\end{enumerate}
These modules interact in a closed-loop fashion, allowing for a self-contained material cycle. Items are retrieved from the Automated High-Bay Warehouse, undergo processing at the Multi Processing Station With Oven, sorted by color using the Sorting Line With Color Detection, and are ultimately returned to the Automated High-Bay Warehouse for storage. The setup can be seen in Figure~\ref{fig:mftb_neutral}.
We write a simple natural language description of the specifications of the HighBay module and we run the corresponding prompt through our pipeline. We deploy the formally verified code on the MFTB to verify its operation on physical hardware. \textbf{The specification, results and demonstration videos are included in our website \footnote{https://sites.google.com/uci.edu/llm4plc/home}.}

\section{Results}
\label{sec:results}
In this section, we present the outcomes of our empirical evaluations. We measure the efficacy of our method -- LLM4PLC -- in terms of three critical metrics: pass rate, compiler error count, and a human evaluation of code quality. We also compare the needed engineering-hours needed to set up each of the solutions. These metrics offer a comprehensive assessment of our approach, covering aspects such as accuracy, robustness, and practical utility in code generation processes.  The experimental setup for each of these metrics is outlined in the following subsections.

\subsection{Pass Rate}
\begin{figure}[h!]
    \centering
    \includegraphics[width=\linewidth]{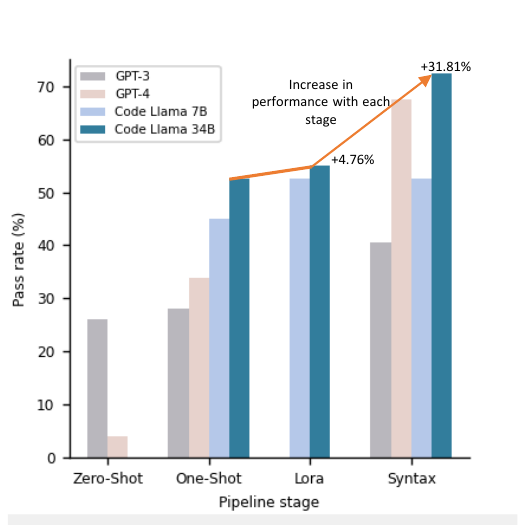}
    \caption{Pass rate for each model and configuration type}
    \label{fig:passrate}
\end{figure}

For evaluating the pass rate, we pass the set of 40 dedicated test files through various configurations of LLM4PLC pipeline stages. For each of the resulting 40 input files, we determine the pass rate as defined in Section \ref{sec:metrics}. The results are presented in 
Figure~\ref{fig:passrate},

where \textit{Zero-Shot} has results for the Zero-Shot prompt, \textit{One-Shot} has results for the One-Shot prompt that includes SCL code as an example, \textit{LoRA} has results for the LoRA-Finetuned Models and One-Shot prompt, and \textit{syntax} incorporates the grammar checker in addition to the LoRA and the One-Shot prompt. As more components are added to our pipeline as we progress through the stages, the compilation pass rate increases, reflecting the efficacy of our proposed method, specifically in the final stage, which includes the One-Shot prompt, a LoRA, and a grammar checker. Evidently, each added component makes a positive contribution to the pass rate. The most significant increase can be seen when the grammar checker is incorporated, which when used with the optimized prompt and LoRA, results in a 72.5\% pass rate for the Code Llama 34B model.
\subsection{Compiler Error Count}
\begin{figure}[h!]
    \centering
    \includegraphics[width=\linewidth]{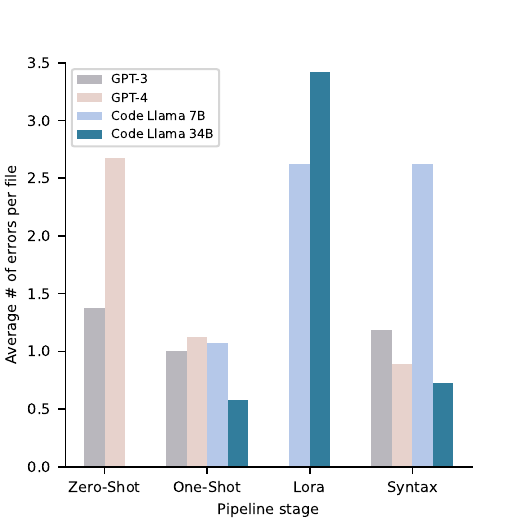}
    \caption{Cumulative syntax errors recorded over the dataset for each model configuration}
    \label{fig:errors}
\end{figure}

Next, we compute the number of compiler errors generated for each output code file as per the procedure defined in \ref{sec:metrics}. We then average the number of errors over the set of test files in order to normalize the results. The average errors per test file is then the final metric we obtain. The results are presented in Figure~\ref{fig:errors}
.
The errors for \textit{LoRA} and \textit{syntax} are larger than that of just the \textit{One-Shot}. However, as evidenced by the Pass Rate over each successive configuration, our pipeline is able to correct these errors such that the compilation rates are higher for \textit{LoRA} and \textit{syntax} compared to just \textit{One-Shot}. One pattern we observed is the tendency for the LLM in this setting to expand variable definition blocks well beyond their intended size. These blocks often contained multiple syntax errors per line, and combined with the excessive generation of them in some files led to the abnormally high compiler error rate for this configuration. Note that the naive prompt generation for the Code Llama models had a $0\%$ pass rate its error rate was omitted from the graph for that reason.

\subsection{Human Quality Metrics}
The human quality metrics are assessed using a panel of experts of various experience with PLC programming\footnote{Our research team obtained an approval from the institutional IRB through an exemption given the academic nature of our questionaire}. The panel evaluates the generated code based on predefined criteria: correctness, maintainability, and conformance to industry coding standards. The experts were informed of the purpose of this study but they were not informed which model created which code. We asked the participating individuals to score the criteria by answering the following questions:
\begin{itemize}
\item Correctness: On a scale of 1-10, how accurately does the generated code perform the intended function without errors?
\item Maintainability: On a scale of 1-10, how easy is it to understand, modify, and extend the code?
\item Conformance to Industry Coding Standards: On a scale of 1-10, how well does the code adhere to established best practices and coding standards?
\end{itemize}
This qualitative evaluation complements our quantitative metrics, providing a multi-faceted view of the system's performance. High scores in this area would signify that the generated code is not only accurate and robust but also practical for real-world applications. The results are presented in Table~\ref{tab:score}. Surprisingly, even though our pipeline did not take the prevalent best practices into consideration, the use of LLM4PLC enhanced the perceived quality of the code. 

\begin{table}[h!]
    \centering
    \begin{tabular}{|p{1.5cm}||c|c|c|}
    \hline
        Model             & Correctness   & Maintainability         & Best Practices\\ \hline
        LLama-34B Naive   & 2.25          & 3.25                    &2.5 \\\hline
        LLama-34B LLM4PLC & 6.5           & 4.75                    &4.0 \\\hline
        GPT-4 Naive       & 2.25          & 3.75                    &2.75 \\\hline
        GPT-4 LLM4PLC     & \textbf{7.75} & \textbf{6.125}          & \textbf{6.0}\\\hline
    \end{tabular}
    \caption{Average Expert-Appointed score}
    \label{tab:score}
\end{table}

\subsection{Engineering Effort}
The question of engineering effort is critical when choosing between GPT-based solutions, LoRA-augmented LLMs, and traditional hand-programming techniques for PLCs. Our empirical studies show marked differences in the time required for each approach, offering insights into their practical applicability.

For GPT models, the setup is remarkably efficient, often taking only a matter of minutes to integrate the API and commence code generation. In contrast, setting up a LoRA for an existing open-source model like Code Llama entails a considerable amount of time to collect data and train.

Hand-programming of PLCs, while a well-understood method, is by far the most time-consuming, often requiring orders of magnitude more time than the iterative LLM-based approaches we present. While this method provides the highest degree of control and specificity, it also demands the most significant investment in terms of time and specialized human resources. For this metric, we take the average time it took for each member of our team to setup the program for the high-bay module.

We present the table of effort needed in Table~\ref{tab:effort}.

\begin{table}[h!]
    \centering
    \begin{tabular}{|c|c|c|}\hline
         Approach      &  Setup Time& Run-Time\\ \hline\hline
         Hand-Coding   &  -         &  12 hours\\\hline
         Lora-Based    &  32 Hours  &  6 minutes\\\hline
         Out-Of-The-Box&  -         & \textbf{2 minutes} \\\hline
    \end{tabular}
    \caption{Engineering-hours required for each approach}
    \label{tab:effort}
\end{table}

\vspace{-1.5em}

\vspace{-0.5em}
\section{Discussion}
\label{sec:discussion}
This section delves into the broader implications of our proposed pipeline, LLM4PLC. We explore its conceptual and practical impacts. Then, we examine how the pipeline accelerates the development iteration cycle, thereby allowing for more efficient Industrial Automation workflows. We include an excerpt from an example of the MBD Plan generated for the MFTB High Bay in Listing~\ref{lst:mbd}, an example of an erroneous generation in Listing~\ref{lst:bad}, and its corresponding correction after the nuXmv and Compiler Checks in Listing~\ref{lst:good}

\begin{lstlisting}[label=lst:mbd, caption=Excerpt of a generated MBD Transition Logic]
    LIST THE TRANSITIONS BETWEEN STATES:
    STATE 0 -> STATE 1: IF "RED BTN" = TRUE AND NO COMPONENT IN HAND
    STATE 1 -> STATE 2: IF xpos = 734 AND ypos = 405
    STATE 2 -> STATE 3: ONCE ARM EXTENDED AND COMPONENT LIFTED
    [...]
\end{lstlisting}

\begin{lstlisting}[label=lst:bad, caption=Excerpt of a generated erroneous Structured Text Code]
componentHomeSlot: TUPLE OF (INT, INT);
	    [...]
IF xPos < 1950 THEN
	    [...]
IF "RED BTN" THEN
   IF xPos < 734 THEN
	    [...]
\end{lstlisting}

\begin{lstlisting}[label=lst:good, caption=Excerpt of a generated corrected Structured Text Code]
componentHomeSlot: ARRAY[1..2] OF INT;
		[...]
IF xPos < 1900 THEN
		[...]
IF "RED BTN" AND NOT componentInHand THEN
   IF xPos < 734 THEN
		[...]
\end{lstlisting}

\subsection{Impact}
Our proposed pipeline accelerates the iteration cycle, from model training to code verification. This acceleration improves not only development speed, but also the quality and reliability of the generated code, thus having a substantial impact on the end application.

Moreover, the pipeline's speed enables real-time or near-real-time code adaptation, making it possible to apply updates or patches in time-constrained environments. This responsiveness is particularly invaluable in critical systems where failure to correct an issue promptly could result in substantial economic losses or safety risks. 

\subsection{GPT vs Code Llama}
In our experiments,  the fine-tuned Code Llama 34B model with the use of a grammar checker outperformed the general-purpose GPT models like GPT-3.5 and GPT-4 on several key metrics. Even with our grammar checker, the GPT models underperform compared to Code Llama 34B. This can largely be attributed to the fact that fine-tuning allows the LLM to learn the nuances of Structured Text, which GPT models have not been exposed to before. Note that due to the closed-source nature of the GPT models, fine-tuning is not a possibility. Moreover, our experiments showcased the efficacy of LoRAs for increasing model performance. It is important to note that an increase in model size (i.e. the total number of parameters) can affect the performance of a model, as a larger number of parameters allows for richer context-awareness, which improves output accuracy. We attribute the better performance of GPT-4 with the grammar checker compared to Code Llama 7B with the LoRA and grammar checker to this reason. Even so, Code Llama 34B with the LoRA and grammar checker, despite having fewer parameters than GPT-4, outperforms GPT-4 with the grammar checker, demonstrating the strength of LoRAs when it comes to improving the quality of an LLM's output on specific output domains.

Furthermore, The closed-source nature of GPT models restricts auditability and customization, creating potential barriers for applications requiring rigorous verification or specialized features. This opacity not only hinders the research community from fully understanding the limitations of these models but also inhibits collective efforts to improve upon these shortcomings. Models that allow fine-tuning can be customized, which evidently can improve performance and generalization to unseen data.
\section{Industry Challenges}
\label{sec:industry}

As automation and digitalization increasingly pervade industrial settings, several emergent challenges must be addressed.

The fast advancement of technology creates major challenges in the industry to maintain a sufficiently skilled workforce that can, on the one hand, leverage new technologies quickly, and on the other, preserve the industrial know-how that has been acquired from years of practical and hands-on experience. The use of LLMs to collect relevant technical knowledge and leverage it for automating engineering tasks and for providing guidance to engineers is therefore of special interest in the industry, as it can accelerate the process of knowledge gain for new staff, reduce the negative effect of knowledge loss from staff rotation, and lessen the influence of staff skillset on engineering efficiency and quality. In addition, the overall engineering cost is expected to be reduced as a consequence of increased automation of the engineering tasks.

Another challenge is the issue of explainability and trust. Industrial applications often have stringent safety and reliability standards. Incorporating machine learning solutions that provide not just high performance but also interpretability will be crucial. The ability to understand and trust the decisions and actions of an AI system is a non-negotiable requirement in critical infrastructures where errors can lead to significant economic or human loss. Although our study shows that the code achieves a good degree of interpretability, un-regulated use of such methods will inevitably lead to unverified and unmoderated deployments in critical systems.

\section{Conclusion}
\label{sec:conclusion}
In summary, our research introduces the LLM4PLC framework, a user-guided iterative pipeline designed to improve the automated code generation capabilities of Large Language Models, specifically for Industrial Control Systems operated by Programmable Logic Controllers. By incorporating user feedback and leveraging external verification tools, we significantly enhance the model's output validity. This pipeline is further refined by Prompt Engineering techniques and model fine-tuning methods like LoRAs. Empirical validation on a FischerTechnik Manufacturing TestBed demonstrates notable improvements in both the rate of successful code generation and the quality of the generated code, as rated by experts. Our contributions pave the way for more reliable and efficient applications of LLMs in industrial settings, inching us closer to achieving verifiably correct program generation in these critical domains.

\section*{Acknowledgments}
This research was supported by Siemens AG. Any opinions, findings, conclusions, or recommendations expressed in this paper are those of the authors and do not necessarily reflect the views of the funding body.

\printbibliography
 
\end{document}